\documentclass[aps,prl,superscriptaddress,amsfonts,amsmath,amssymb,showpacs,floatfix,reprint,longbibliography,footinbib]{revtex4-2}

\usepackage{url}
\usepackage{bm}
\usepackage{graphicx}
\usepackage[percent]{overpic}
\usepackage{amsmath}
\usepackage{amstext}
\usepackage{amssymb}
\usepackage{amsfonts}
\usepackage{amsbsy}
\usepackage{verbatim}
\usepackage{color}
\usepackage[colorlinks=true, urlcolor=blue, linkcolor=blue, citecolor=blue, pdftex]{hyperref}
\usepackage[capitalise]{cleveref}
\usepackage{multirow}
\usepackage{floatrow}
\usepackage{float}
\usepackage{tikz}
\usepackage{gensymb}
\usepackage{textcomp}
\usepackage{enumitem}
\usepackage[version=3]{mhchem}
\usepackage{afterpage}
\usepackage{pbox}
\usepackage{makecell}
\usepackage{dsfont}
\usepackage{siunitx}
\usepackage{stackengine,wasysym,scalerel}
\usepackage{latexsym}
\usepackage{cancel}
\usepackage{array, multirow, bigdelim, makecell, booktabs}
\usepackage[misc]{ifsym}
\usepackage{times}
\usepackage{scrextend}
\usepackage{footmisc}

\usepackage{ulem}

\begin{document}

\title{Quantum Coulomb Liquids of Different Rank in the Breathing Pyrochlore Antiferromagnet}
  
\author{Lasse Gresista}
\altaffiliation[These authors contributed equally to the project.] {}
\affiliation{Institute for Theoretical Physics, University of Cologne, 50937 Cologne, Germany}
\affiliation{Department of Physics, Indian Institute of Technology Madras, Chennai 600036, India}
\author{Daniel Lozano-G\'omez}
\altaffiliation[These authors contributed equally to the project.] {}
\affiliation{Institut f\"ur Theoretische Physik and W\"urzburg-Dresden Cluster of Excellence ctd.qmat, Technische Universit\"at Dresden, 01062 Dresden, Germany}
 \author{Matthias Vojta}
 \affiliation{Institut f\"ur Theoretische Physik and W\"urzburg-Dresden Cluster of Excellence ctd.qmat, Technische Universit\"at Dresden, 01062 Dresden, Germany}
\author{Simon Trebst}
\affiliation{Institute for Theoretical Physics, University of Cologne, 50937 Cologne, Germany}
\author{Yasir Iqbal}
\affiliation{Department of Physics, Indian Institute of Technology Madras, Chennai 600036, India}

  \date{\today}


\begin{abstract}
Emergent gauge fields and Coulomb liquids have long been central to the physics of frustrated pyrochlore magnets, yet their realization beyond conventional, i.e. rank-1 $U(1)$, spin ice and into fully quantum higher-rank regimes has remained elusive. Here we provide a controlled demonstration of this physics in the spin-$\tfrac{1}{2}$ quantum Heisenberg antiferromagnet on the breathing pyrochlore lattice with symmetry-allowed Dzyaloshinskii--Moriya interactions, using the pseudofermion functional renormalization group. We show that tuning the breathing asymmetry stabilizes extended quantum analogues of both rank-1 and rank-2 $U(1)$ Coulomb liquids within a single microscopic model, directly distinguished by their characteristic pinch-point morphologies in momentum space. This provides the first controlled quantum realization in three dimensions where gauge theories of different rank emerge within a single microscopic spin Hamiltonian. In addition, quantum fluctuations qualitatively reshape the classical nearest-neighbor atlas of phases, causing an incommensurate spiral instability and an extended quantum-disordered regime without dipolar order, both absent from the classical model. Our results establish the breathing pyrochlore as a timely and experimentally relevant platform where higher-rank gauge constraints, conventional magnetic order, and fluctuation-driven quantum phases compete on equal footing, opening a direct route to diagnosing emergent gauge structure in three-dimensional quantum magnets.
\end{abstract}

\maketitle


{\it Introduction.---}
Frustrated magnetic lattices provide a natural setting for the emergence of collective phases that evade conventional symmetry-breaking descriptions, including spin liquids governed by emergent gauge fields~\cite{Knolle_Moessner_2019,Yan2024a,Yan2024b,Davier2023,Savary-2017}. The pyrochlore lattice of corner-sharing tetrahedra occupies a central role, as its strong geometric constraints are capable of generating extensive ground-state degeneracies and rendering the system exceptionally sensitive to perturbations~\cite{Rau2019ARCMP,Hallas-AnnRevCMP,lozano-2023,benton2016,Yan-2020,Smith_CeZrO,zhang_scheie2025intrinsicquantumdisorderyb2ti2o7,gresista_quantum_pinch_line,Bramwell-2001,Wong2013}. Over the past decade, these degeneracies have been systematically exploited to classify the phases of the most general bilinear nearest-neighbor Hamiltonian on the regular pyrochlore lattice~\cite{KTC_2024_phase,Wong2013,yan2017}.
In the classical limit, this program has culminated in a comprehensive atlas, comprising multiple magnetically ordered phases~\cite{Rau2019ARCMP,KTC_2024_phase}, nematic phases~\cite{Taillefumier_2017,Francini2024nematicR2,francini2025exactnematicmixedmagnetic}, and an extensive family of Coulomb spin liquids characterized by emergent gauge constraints of different rank~\cite{lozano_2024_atlas}. Notably, the regular pyrochlore lattice supports not only the familiar rank-1 $U(1)$ Coulomb phase~\cite{Moessner-1998,Taillefumier_2017,Bramwell-2001} but also rank-2 Coulomb liquids associated with a higher-rank gauge structure~\cite{gresista_quantum_pinch_line,benton2016,lozano-2023,niggemann2023}, for which the minimal excitations are gauge charges with restricted mobility known as fractons~\cite{Pretko-2017,Pretko-2017b,Prem-2018,niggemann2025classicalfractonspinliquid,niggemann2025gaplessfractonquantumspin,lozano_2024arxiv}. 

Studies of the quantum $S=\tfrac12$ model indicate that many of these phases possess quantum counterparts, while also revealing fluctuation-driven regimes absent from the classical limit~\cite{gresista_quantum_pinch_line,zhang_scheie2025intrinsicquantumdisorderyb2ti2o7,lozano-2023}. As a result, nearest-neighbor exchange models on the regular pyrochlore lattice are now comparatively well understood. A central open question emerging from this progress is whether higher-rank Coulomb phases -— thus far firmly established only in classical settings -— can survive quantum fluctuations in three-dimensional magnets. This question has gained renewed urgency in light of recent theoretical and experimental activity on higher-rank Coulomb phases, which has sharpened the distinction between rank-1 and rank-2 gauge structures while leaving their stability in fully quantum three-dimensional magnets unresolved.

The breathing pyrochlore lattice, in which ``up'' and ``down'' tetrahedra are inequivalent~\cite{Ghosh-2019b,Iqbal-2019}, provides a minimal and experimentally relevant setting in which to address this question. The reduced symmetry enlarges the interaction parameter space and modifies the local constraints in a way that fundamentally alters the balance between classical degeneracy and quantum fluctuations. While previous works have identified selected ordered phases~\cite{Ghosh-2019b}, regimes dominated by single-tetrahedron physics~\cite{Dissanayake2022}, and both classical~\cite{Yan-2020} and quantum~\cite{Iqbal-2019} spin-liquid behavior, a unified understanding of how gauge structure, magnetic order, and quantum fluctuations compete on the breathing pyrochlore lattice remains lacking.

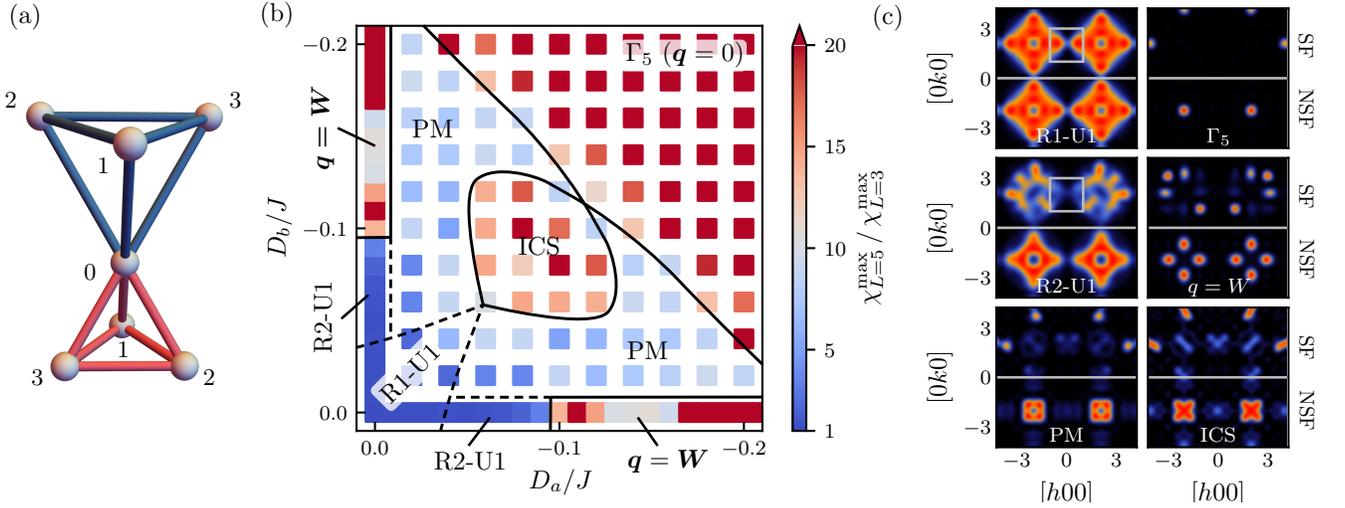
\begin{figure*}[ht!]
    \centering
     \begin{overpic}[width=.98\textwidth]{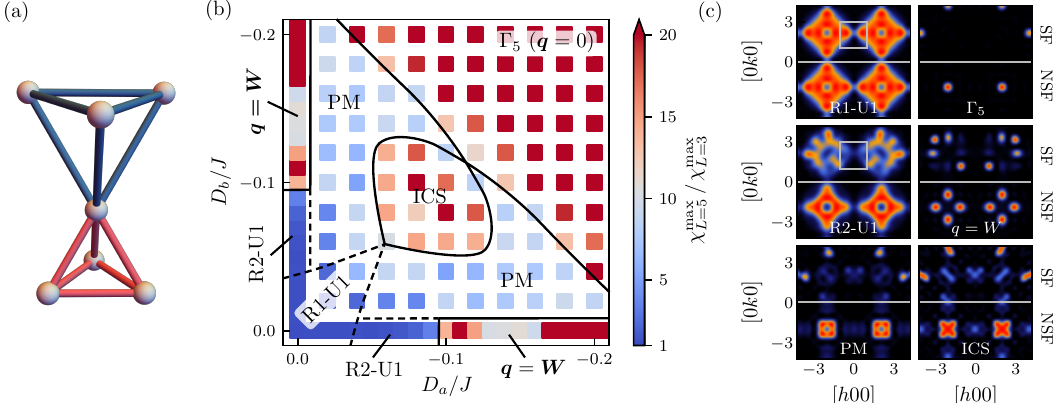}
     \put(0,30){$2$}
     \put(17,30){$3$}
     \put(6,17){$0$}
     \put(7,25){$1$}
     \put(15,9){$2$}
     \put(2,9){$3$}
     \put(8.5,11.1){$1$}
     \end{overpic}
   \caption{\textbf{Quantum phase diagram.} 
(a) Unit cell of the breathing pyrochlore lattice with the sublattice labels indicated near every site. 
(b) Quantum phase diagram obtained from pf-FRG. Solid lines indicate the boundaries between regions with dipolar magnetic order (red) and quantum paramagnetic (PM) regions (blue) without dipolar order. Dashed lines serve as guides to the eye and indicate regimes where the structure factor within the quantum PM regions changes. 
(c) Polarized neutron scattering structure factors in the spin-flip (SF, top) and non-spin-flip (NSF, bottom) channels in the $hk0$ plane for each phase, calculated in the low-cutoff limit (see Supplemental Material for the corresponding structure factor in the $hhl$ plane).}
    \label{fig:quantum-phase-diagram}
\end{figure*}

In this work, we address this gap by studying the classical and quantum $S=\tfrac12$ Heisenberg antiferromagnet on the breathing pyrochlore lattice with symmetry-allowed Dzyaloshinskii--Moriya (DM) interactions,
\begin{eqnarray}
\mathcal{H}&=&J\sum_{\langle ij\rangle } \bm{S}_i\cdot \bm{S}_j
+D_a\sum_{\langle ij\rangle_A } \bm{d}_{ij}\cdot( \bm{S}_i\times\bm{S}_j)\nonumber\\
&&+D_b\sum_{\langle ij\rangle_B } \bm{d}_{ij}\cdot( \bm{S}_i\times\bm{S}_j)
=\sum_{\langle ij\rangle }S^{\alpha}_i J^{\alpha\beta}_{ij}S^{\beta}_j,
\label{eq:H+DM}
\end{eqnarray}
focusing on the experimentally relevant indirect DM regime~\cite{Elhajal_2005_a}, see \textit{End Matter} for a definition of the $\bm d_{ij}$ vectors. For simplicity, we assume equal $J>0$ on both types of tetrahedra and consider $D_{a,b}\leq0$, as this was identified~\cite{Yan-2020} as the most interesting parameter regime at the classical level~\cite{fn_rank2width}. 
We demonstrate that this minimal model already stabilizes robust quantum analogues of both rank-1 and rank-2 Coulomb spin liquids, directly distinguished by their momentum-space correlation signatures, and that quantum fluctuations generate an incommensurate spiral instability and an extended non-dipolar quantum-disordered regime, neither of which is realized in the classical model, see Fig.~\ref{fig:quantum-phase-diagram}. To place these quantum results on firm footing, we first establish the structure of the classical phase diagram and its associated correlation fingerprints, which then serve as a reference for the quantum analysis subsequently presented.

{\it Classical model.---}
We begin by analyzing the classical limit of Eq.~\eqref{eq:H+DM}, which provides a necessary baseline for disentangling kinematic constraints, emergent gauge structure, and fluctuation-induced ordering tendencies. In particular, the classical model exposes which features of the breathing pyrochlore phase diagram are dictated by exact degeneracies and flat bands, and which arise only once quantum fluctuations are introduced. Thus far, the classical breathing pyrochlore model has been explored only along two special manifolds in parameter space: the line $D_a=D_b$ (corresponding to the regular pyrochlore lattice)~\cite{Noculak-2023,Elhajal_2005_a,Canals_Elhajal_2008}, and the line $D_b=0$ (or equivalently $D_a=0$). In the regular pyrochlore lattice, Ref.~\cite{Noculak-2023} found that a $\Gamma_5$ phase is selected at low temperatures by an order-by-disorder mechanism whenever $D_a <0$. In the limiting case where $D_a=D_b=0$, i.e., at the Heisenberg point, a rank-1 $U(1)$ classical spin liquid is realized~\cite{Moessner-1998}. For the $D_b=0$ line, Ref.~\cite{Yan-2020} found that a rank-2 $U(1)$ classical spin liquid, constrained by an emergent (energy) Gauss's law $\partial_{i}E_{ij}=0$~\cite{Yan-2020,lozano_2024_atlas}, where
(in terms of coarse--grained fields $m_X$ which transform as irreducible representations $X$ of the point-group symmetry $T_d$)
\begin{eqnarray}
    E=
       \begin{pmatrix}
	2 m_E^2 &  \sqrt{3}m_{T_{1-}}^z &   \sqrt{3}m_{T_{1-}}^y \\
	 \sqrt{3}m_{T_{1-}}^z & -\sqrt{3} m_{E}^1 - m_E^2 &   \sqrt{3}m_{T_{1-}}^x \\
	\sqrt{3}m_{T_{1-}}^y &  \sqrt{3} m_{T_{1-}}^x & \sqrt{3} m_E^1 - m_E^2
	\end{pmatrix},
\end{eqnarray}
is realized at \textit{intermediate} temperature before giving way to a long-range ordered state with $\bm q=W$ ordering wave vector at low temperatures. 

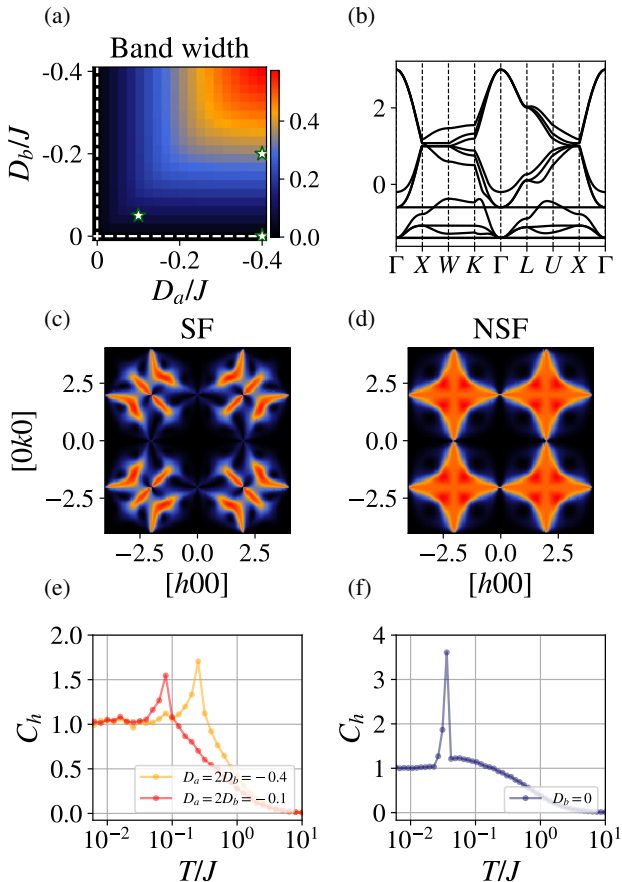
\begin{figure}[t]
    \centering
      \begin{overpic}[width=\textwidth]{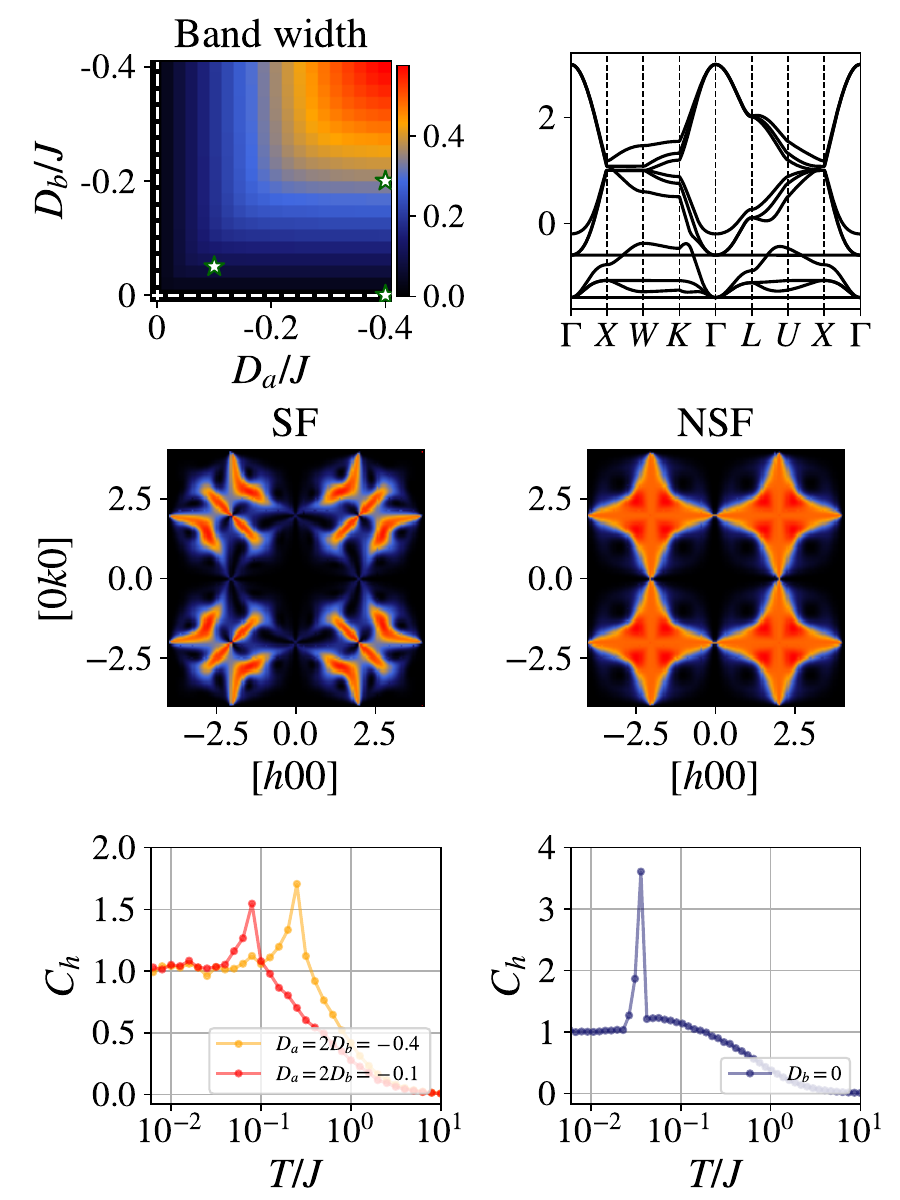}
        \put(7,100){(a)}
        \put(42,100){(b)}
        \put(7,65){(c)}
        \put(42,65){(d)}
         \put(7,34){(e)}
        \put(42,34){(f)}
    \end{overpic}
    \caption{\textbf{Classical phase diagram.} (a) Band width of the lowest-energy band as a function of the DM couplings $D_a,D_b$. (b) Interaction matrix $J^{\alpha\beta}(\bm q)$ energy bands for a set of parameters where $D_b=0$. (c,d) Low-temperature prediction of the spin-flip (SF) and non-spin-flip (NSF) channels of the polarized neutron scattering obtained via the self-consistent Gaussian approximation (SCGA) for a set of parameters where $D_b=0$. (e,f) Specific heat obtained via classical MC simulations for three representative points marked by stars in panel (a). For two of them the system orders into a $\Gamma_5$ phase and for one the low-temperature phase is a $\bm q=\bm W$ phase. }
    \label{fig:classical_pinch-points}
\end{figure}

We now consider the more general case where $D_a, D_b\neq 0$. As the two types of tetrahedra, i.e.\ the ``up'' and ``down'' tetrahedra shown in Fig.~\ref{fig:quantum-phase-diagram}(a), are related to each other by a $C_2$ rotation~\cite{Rau2019ARCMP}, the phase diagram is symmetric with respect to the $D_a=D_b$ line. This symmetry is reflected in the evolution of the spectrum of the interaction matrix $J^{\alpha\beta}_{\mu\nu}(\bm q)$ and, in particular, in the bandwidth of the lowest-energy band as shown in Fig.~\ref{fig:classical_pinch-points}(a). From this spectrum, the possible magnetic phases can be identified via the wave vector $\bm q$ at which the minimum energy is reached. In this diagram, the white line labels the $D_a=0$ line, where the bandwidth vanishes and two low-energy flat bands are generated, see Fig.~\ref{fig:classical_pinch-points}(b). These two low-energy flat bands lead to the realization of the rank-2 U(1) spin liquid studied in Ref.~\cite{Yan-2020}, whose higher-rank nature is exposed by the observation of two- and fourfold pinch points in the spin correlation functions
illustrated in Fig.~\ref{fig:classical_pinch-points}(c,d) obtained via the self-consistent Gaussian Approximation (SCGA)~\cite{SCGA_Canals_kagome,SCGA_Canals_pyrochlore,lozano_2024_atlas}. Away from the $D_a=0$ line, the lowest-energy bands become dispersive and acquire a minimum value at the $\Gamma$ point and along the $\Gamma$-$\rm L$ line; see {\it End Matter}. 

Beyond the band-spectra analysis, we perform classical Monte Carlo (MC) simulations in two regions: one along the $D_a=0$ line, and at generic points where $D_a\neq D_b$. In every case, our simulations detect a symmetry-breaking phase transition at low temperatures, see Fig.~\ref{fig:classical_pinch-points}(e,f). Along the $D_a=0$ line, and in agreement with Ref.~\cite{Yan-2020,pollet_ML_R2}, our simulations expose a $\bm q\neq 0$ low-temperature phase where magnetic Bragg peaks are observed at the $W$ points, i.e. $\bm q\in\{[111],[113],[331]\}$ and symmetry-related directions, see {\it End Matter}. For generic $D_a, D_b < 0$, our simulations detect a low-temperature $\bm q=0$ phase, corresponding to the so-called $\Gamma_5$ phase~\cite{Hallas-AnnRevCMP,Noculak-2023,Sarkis_YbGeO}. In the low-temperature limit ($T\ll T_c$), this phase corresponds to the so-called $\psi_2$ order, where the spin degrees of freedom orient along the local $x$-axis. Performing a more precise identification of the magnetic phases in the $D_a$-$D_b$ plane would require a more extensive numerical analysis 
as the one done in Ref.~\cite{Noculak-2023}, which we leave for future work. For more information regarding our classical MC simulations, we refer the reader to {\it End Matter}.

{\it Quantum model.---}
We now turn to the quantum $S=1/2$ model and map out its phase diagram in the $D_a-D_b$ plane, focusing on the role of quantum fluctuations in the $T \to 0$ limit. Our aim is to assess how these fluctuations modify the classical phase structure, including whether they give rise to new phases and whether characteristic signatures of the classical rank-2 $U(1)$ spin liquid persist in the quantum model. To this end, we employ the pseudofermion functional renormalization group (pf-FRG) method~\cite{Muller-2024}. The pf-FRG is particularly well suited here as it treats quantum fluctuations and competing ordering tendencies on equal footing without bias toward any specific order parameter.

Within the pf-FRG framework, the static ($\omega = 0$) spin-spin susceptibility is computed as function of an infrared cutoff $\Lambda$,
\begin{equation}
\label{eq:correlations}
\chi_{ij}^{\Lambda, a b} =  
\int_0^\infty d\tau \,
\left\langle
\hat{T}_\tau \hat{S}^a_i(\tau)\hat{S}^b_j(0)
\right\rangle^{\Lambda},
\end{equation}
where $\hat{T}_\tau$ denotes the imaginary-time ordering operator. The renormalization-group flow is integrated from the trivial limit $\Lambda \to \infty$ down to $\Lambda \to 0$, where physical observables are obtained. Throughout the integration, spatial correlations are retained up to a maximum bond distance of $L=5$, corresponding to 361 correlated lattice sites, while longer-range correlators are neglected. Dipolar magnetic order is signaled by a divergence (“flow breakdown”) in selected components of $\chi^{\Lambda,ab}$ at characteristic wave vectors $\bm{k}_{\mathrm{max}}$ and at a finite critical scale $\Lambda_c$. By contrast, the absence of such a breakdown indicates a quantum paramagnetic state. The momentum-resolved susceptibility and the neutron-scattering structure factor in the low-cutoff limit further allow us to characterize the nature of the ground state. This allows us to compute the $T = 0$ quantum phase diagram for $D_a, D_b < 0$, shown in Fig.~\ref{fig:quantum-phase-diagram}(b), which we discuss in the following. Additional technical details are provided in the {\it End Matter}.

\begin{figure}[ht!]
    \centering
    \includegraphics{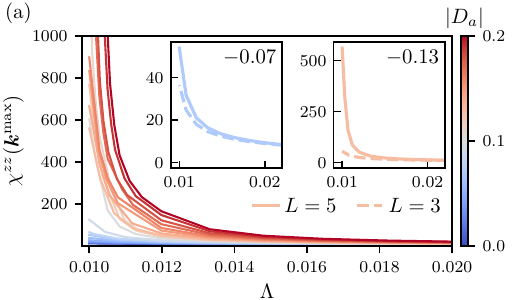}
    \includegraphics{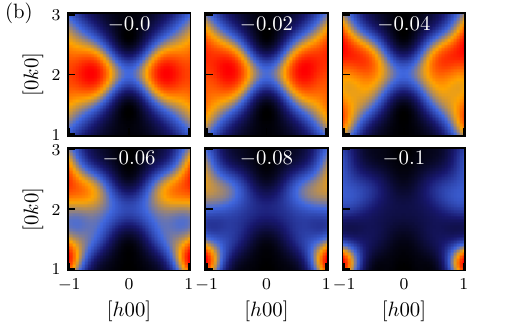}
    \caption{\textbf{Flows and pinch-points for $\mathbf{D_b = 0}$.} 
    (a) Flow of $\chi^{zz}(\bm{k}^\mathrm{max})$ for increasing $D_a$ at lattice truncation $L=5$. Both the absolute magnitude and the system-size scaling increase sharply at $D_a = -0.1$ in the low-cutoff limit, which we interpret as the transition into the $\bm{q}=W$ ordered phase. The insets show the corresponding flows for $L = 3$ and $L = 5$ in the quantum paramagnetic ($D_a = -0.07$) and the $\bm{q}=W$ $(D_b = -0.13)$ phases, highlighting the qualitative difference in their scaling behavior.
    (b) Polarized neutron scattering structure factors in the spin-flip channel, zoomed into the region around $[0,0,2]$ as indicated in Fig.~\ref{fig:quantum-phase-diagram}(c), for different $D_b$ within the spin-liquid phases. The emerging four-fold pinch point is a characteristic signature of the rank-2 $U(1)$ spin liquid.
    }
    \label{fig:quantum pinch-points}
\end{figure}

We first focus on the line $D_b = 0$ with $D_a < 0$. Along this line, the classical model exhibits successive transitions upon cooling from a rank-1 $U(1)$ spin liquid to a rank-2 $U(1)$ spin liquid and, finally, to $\bm{q}=\bm{W}$ order \cite{Yan-2020}. In contrast, for small but finite $|D_a|$, the quantum model shows no indication of a transition into a magnetically ordered phase. Instead, it exhibits a smooth susceptibility flow that is largely independent of the spatial cutoff $L$ and shows no sign of a flow breakdown, see Fig.~\ref{fig:quantum pinch-points}(a). Moreover, the neutron-scattering structure factor around the [220] momenta displays a clear twofold pinch-point structure, consistent with correlations characteristic of the rank-1 $U(1)$ spin liquid found in the classical model, see Fig.~\ref{fig:quantum pinch-points}(b).

Upon further increasing $|D_a|$, around $D_a \simeq -0.04J$, the susceptibility flow remains free of a clear breakdown while the twofold pinch points gradually become asymmetric and evolve into a fourfold pinch-point structure. This is consistent with a quantum rank-2 U(1) Coulomb liquid characterized by fourfold pinch-point correlations in this region of the phase diagram. For $D_a \lesssim -0.1J$, the pinch points disappear and the structure factor instead develops pronounced peaks at momenta symmetry-related to $\bm{k}^{\mathrm{max}}=\bm{W}$, closely resembling the classical structure factors in the corresponding ordered phase [cf. Fig.~\ref{fig:quantum-phase-diagram}(c) and Fig.~\ref{fig:classical_pinch-points}]. Concomitantly, we observe a sharp increase in both the absolute magnitude and the truncation-length dependence of $\chi^{zz}(\bm{k}^{\mathrm{max}})$ at low cutoff scales [Fig.~\ref{fig:quantum pinch-points}(a)]. We identify this behavior as the onset of $\bm{q}=\bm{W}$ order at small cutoff scales ($\Lambda \approx 0.01J$), in close analogy to the classical transition, which occurs at comparably low temperatures $T_c \lesssim 0.01J$. 

Considering the more general case with $D_a,D_b \neq 0$, the fourfold pinch-point structure -- and thus the putative rank-2 $U(1)$ spin-liquid phase -- is stable only along the lines $D_b=0$ (or, equivalently, $D_a=0$), and is absent away from this limit. By contrast, twofold pinch points persist over a finite region around the Heisenberg point, indicating a robust rank-1 $U(1)$ spin-liquid phase [see dashed lines in Fig.~\ref{fig:quantum-phase-diagram}(b)]. Along the diagonal $D_a=D_b$ (corresponding to the regular pyrochlore lattice with DM interactions), the rank-1 $U(1)$ spin liquid undergoes a transition at $D_a=D_b \approx -0.07J$ into an dipolar ordered phase characterized by structure-factor peaks at incommensurate (ICS) momenta within the $hhl$ plane. For even stronger DM interactions $D_a = D_b \leq -0.12 J$, the structure factor then agrees with the $\bm{q}=0$ $\Gamma_5$ phase also found in the classical model for $D_a, D_b < 0$
\footnote{We note that an earlier pf-FRG analysis~\cite{Noculak-2023} did not find the ICS phase and instead a direct transition into the $\Gamma_5$ phase at significantly higher $D_a = D_b \approx -0.4J$. We attribute this difference to the improved frequency treatment and numerical integration scheme employed in the present pf-FRG implementation, which enable a more reliable resolution of low-energy instabilities at scales $\Lambda \lesssim 0.02J$ compared to Ref.~\cite{Noculak-2023}. In addition, we employ a more stringent criterion for identifying the flow breakdown. Further details are discussed in the \textit{End Matter}.
}.

\begin{figure}[t!]
    \centering
    \includegraphics[width=\textwidth]{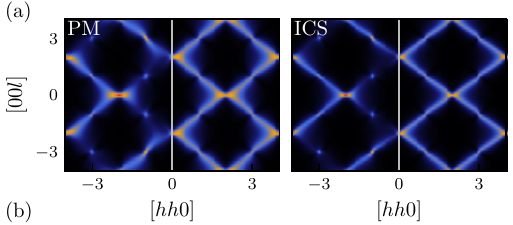}
    \includegraphics{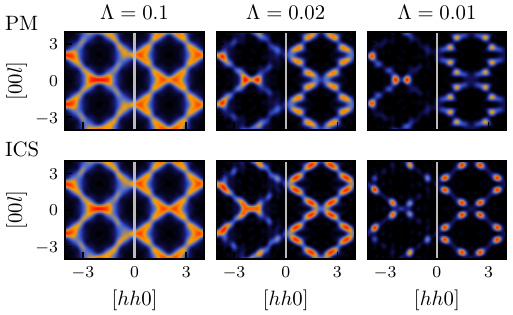}
    \caption{\textbf{Structure factors in the PM and ICS phase in the hhl-plane.} 
    (a) Neutron-scattering structure factor obtained from SCGA in the cooperative paramagnetic regimes at $T > T_c$ for characteristic points where the pf-FRG identifies the PM and ICS phases. The left (right) side show the spin-flip (SF) [non-spin-flip (NSF)] channel. 
    (b) Corresponding neutron-scattering structure factor obtained from pf-FRG at different RG cutoff scales $\Lambda$ (see Supplemental Material for the corresponding structure factor in the $hhl$ plane).}
    \label{fig:evolution_Sq_Lambda}
\end{figure}

Away from the diagonal, $D_a\neq D_b$, in addition to the putative rank-1 and rank-2 $U(1)$ spin liquids, we identify another extended quantum paramagnetic (PM) region that shows no indication of a flow breakdown and lacks pronounced twofold or fourfold pinch points in the structure factor. This phase is labeled PM in Fig.~\ref{fig:quantum-phase-diagram}(b). The neutron-scattering structure factor in the PM phase exhibits comparatively broad, ring-like features around the [220] point in the NSF channel, see Fig.~\ref{fig:quantum-phase-diagram}(c). The absence of dipolar magnetic order in this region suggests that it corresponds to a different type of quantum spin liquid or to a phase characterized by multipolar order, such as spin-nematic or valence-bond order~\cite{gresista_quantum_pinch_line,Francini2024nematicR2,francini2025exactnematicmixedmagnetic}. A direct identification of these phases, however, is beyond the scope of the present pf-FRG approach, as the corresponding susceptibilities would require access to four-spin correlation functions not available within the current framework.

{\it Comparison between the classical and quantum phase diagram.---}
In both the classical and quantum models, we identify rank-1 and rank-2 $U(1)$ spin-liquid phases as well as a conventionally ordered $\bm{q}=0$ phase corresponding to the $\Gamma_5$ manifold, see Figs.~\ref{fig:classical_pinch-points}(a) and \ref{fig:quantum-phase-diagram}(b). In the quantum model, however, two additional phases appear: a dipolar ordered incommensurate phase (ICS) and a non-conventionally ordered paramagnetic phase (PM)
~\footnote{It is worth noting that, for the classical model on the regular pyrochlore lattice with only nearest-neighbor interactions, all magnetically ordered phases identified to date correspond to commensurate $\bm{q} = 0$ magnetic orders~\cite{yan2017,Wong2013,KTC_2024_phase}.}. 
The classical model supports $\bm{q}\neq0$ order only along the $D_b=0$ line. A comparison of the classical and $T=0$ quantum phase diagrams therefore indicates that both the PM and, interestingly, the ICS phase are driven by quantum fluctuations. The emergence of the incommensurate spiral phase highlights a qualitatively new role of quantum fluctuations: rather than merely lifting classical degeneracies, they {\sl dynamically reshape} the momentum-space structure of correlations and generate ordering tendencies absent at the classical level.

Despite the absence of these phases in the classical $T = 0$ phase diagram, the classical correlation functions in the cooperative paramagnetic regime ($T>T_c$) show notable similarities to their quantum counterparts at intermediate RG cutoffs. 
This is illustrated in Fig.~\ref{fig:evolution_Sq_Lambda}, which compares classical structure factors at temperatures slightly above $T_c$ with quantum pf-FRG results for varying cutoff $\Lambda$ at representative points in the PM and ICS regions. In all cases, the spectral weight is concentrated near the same characteristic wave vectors, in particular around the $[220]$, $[200]$, $[420]$, and symmetry-related points. At large cutoffs, the quantum structure factors closely resemble those of the classical cooperative paramagnet and show no qualitative distinction between the PM and ICS regimes. Upon lowering $\Lambda$, however, the spectral weight redistributes in a phase-dependent manner, distinguishing the PM and ICS phases. Since the RG cutoff $\Lambda$ may be loosely interpreted as effective temperature, this behavior suggests a crossover from a classical cooperative paramagnetic regime at high temperatures to the quantum PM or ICS phases at low energies. A similar classical-to-quantum crossover was previously observed~\cite{Itamar_thesis} for a breathing pyrochlore model with further-neighbor interactions, where the quantum correlation functions at elevated temperatures closely resembled their classical counterparts and evolved smoothly toward the low-temperature pf-FRG results upon cooling. Taken together, these results indicate that quantum fluctuations stabilize and ultimately select the ICS or PM phase whereas thermal fluctuations instead select $\Gamma_5$ order.


\textit{Discussion and Outlook.---}
Our results establish the breathing pyrochlore as a minimal and highly tunable platform in which distinct emergent gauge structures and competing orders coexist already at the level of nearest-neighbor Heisenberg and Dzyaloshinskii–Moriya interactions. In the classical limit, the model interpolates between the well-known rank-1 $U(1)$ Coulomb phase of the regular pyrochlore and a breathing-specific rank-2 $U(1)$ regime on the $D_b=0$ line, characterized by flat low-energy bands and higher-rank constraints. Our pf-FRG results demonstrate that sizable regions of this classical landscape survive in the quantum $S=\tfrac12$ model, giving rise to extended quantum analogues of both rank-1 and rank-2 $U(1)$ liquids, while quantum fluctuations additionally stabilize an incommensurate spiral (ICS) phase and a broad quantum-paramagnetic (PM) regime absent from the classical phase diagram.

Our results suggest that the additional breathing degree of freedom can be used to tune the rank of the quantum spin liquid realized, where a rank-1 spin liquid is stabilized for small non-vanishing couplings $D_a,D_b$, while a rank-2 spin liquid is generated in the cases where one of these vanishes. Within our numerical resolution, the rank-2 $U(1)$ regime appears sharply confined to the symmetry lines $D_a = 0$ or $D_b = 0$, where two low-energy flat bands are protected by lattice symmetries. However, given the underlying classical degeneracy and flat-band structure, and the known tendency of quantum fluctuations to stabilize extended disordered phases, we consider it likely that weak symmetry-breaking perturbations $|D_a - D_b| \ll J$ leave remnants of the rank-2 physics over a narrow but finite parameter range beyond the strict symmetry lines. The breathing degree of freedom provides a natural microscopic tuning parameter -- via chemical pressure, hydrostatic pressure, or controlled anisotropies -- for navigating between distinct gauge regimes. While real breathing pyrochlores generally exhibit unequal exchange couplings on the two tetrahedra ($J_a \neq J_b$), the flat-band structure underlying the rank-2 regime persists for unequal Heisenberg couplings; see footnote~\cite{fn_rank2width}. The stability of the higher-rank regime under realistic exchange asymmetry therefore constitutes an important direction for future work. Nevertheless, our results demonstrate that symmetry-allowed nearest-neighbor interactions alone already suffice to generate competing rank-1 and rank-2 gauge structures within a single microscopic setting. 

The extended PM regime identified here, which lacks pinch-point signatures and does not exhibit dipolar order within pf-FRG, constitutes a particularly sharp target for experiments on breathing pyrochlores searching for unconventional quantum-disordered behavior beyond classical Coulomb phases.

Several open directions follow naturally. First, the microscopic nature of the PM regime remains to be resolved: while pf-FRG excludes dipolar magnetic order, it cannot distinguish between a genuine quantum spin liquid and multipolar or valence-bond order, which require access to higher-order spin correlations. Second, the emergence of the ICS phase -- absent in the classical nearest-neighbor model -- points to a fluctuation-driven instability whose origin can be traced to the evolution of the interaction-matrix band minima and the scale-dependent buildup of correlations, highlighting a concrete classical-to-quantum crossover mechanism. Finally, a more general survey of the distinct symmetry-allowed interactions for the breathing pyrochlore, which could result in the identification of novel spin liquid and symmetry-breaking magnetic phases.
Taken together, the breathing pyrochlore emerges as a uniquely clean setting where higher-rank gauge constraints, conventional $\Gamma_5$ order, and fluctuation-stabilized quantum phases compete on equal footing. This makes it an ideal laboratory for diagnosing emergent gauge structure in three-dimensional quantum magnets through directly measurable signatures in momentum space.\\


\vfill
\noindent{\bf \raggedright Data availability} \\
The numerical data shown in the figures and the raw FRG data is available on Zenodo~\cite{pyrochlore_zenodo}.

\clearpage


{\it Acknowledgements.---}
The Cologne group acknowledges partial funding from the DFG within SFB 1238 (project-id 277146847), projects C02, C03. 
D.L.-G. and MV acknowledge financial support from the DFG through the W\"urzburg-Dresden Cluster of Excellence on Complexity,
Topology and Dynamics in Quantum Matter -- \textit{ctd.qmat} (EXC 2147, project-id 390858490) and
through SFB 1143 (project-id 247310070). D.L.-G. is supported by the Hallwachs-R\"ontgen Postdoc Program of ctd.qmat.
The work Y.I. and S.T. was performed in part at the Aspen Center for Physics, which is supported by a grant from the Simons Foundation (1161654, Troyer). This research was also supported in part by grant NSF PHY-2309135 to the Kavli Institute for Theoretical Physics and by the International Centre for Theoretical Sciences (ICTS) for participating in the Discussion Meeting - Fractionalized Quantum Matter (code: ICTS/DMFQM2025/07). L.G. thanks IIT Madras for a Visiting Graduate Student position under the IoE program during which this project was initiated. Y.I. acknowledges support from the Abdus Salam International Centre for Theoretical Physics through the Associates Programme, from the Simons Foundation through Grant No.~284558FY19, from IIT Madras through the Institute of Eminence program for establishing QuCenDiEM (Project No. SP22231244CPETWOQCDHOC).
The numerical simulations were performed on the JUWELS cluster (Forschungszentrum J\"ulich)
and the Noctua2 cluster at the Paderborn Center for Parallel Computing (PC$^2$).
Y.~I.~acknowledges the use of computing resources at HPCE,  IIT Madras.

\section{End Matter}

{\it Definition of DM vectors.---} 
Let us start by providing a definition of the four sublattice site vectors and the six direct DM vectors defined for each bond of a single tetrahedron. For the up tetrahedra, the sublattice site vectors are defined as 
\begin{align}
    \bm r_{0}&=(0,0,0), &
    \bm r_{2}&=\frac{a}{4}(1,0,1), \\
    \bm r_{1}&=\frac{a}{4}(1,1,0), &
    \bm r_{3}&=\frac{a}{4}(0,1,1),\nonumber
\end{align} 
where $a$ is the length of the FCC unit cell, and the sub-indices label the sublattice structure as defined in Fig.~\ref{fig:quantum pinch-points}. Furthermore, and following the notation used by Refs.~\cite{Elhajal_2005_a,Canals_Elhajal_2008,Noculak-2023}, we define direct DM vectors as
\begin{align}
    \bm d_{01}&=(\bar{1},1,0),	&	    \bm d_{12}&=(0,1,1), \\
    \bm d_{02}&=(1,0,\bar{1}),	&	    \bm d_{13}&=(\bar{1},0,\bar{1}), \\
    \bm d_{03}&=(0,\bar{1},1), 	&  	    \bm d_{23}&=(1,1,0)
    \nonumber,
\end{align}
where a DM vector is defined for each bond of the single tetrahedron, in an identical manner for both up and down tetrahedra.
\\

{\it Definition of spin-correlation functions.---}
The most general spin-spin correlation function in the (breathing) pyrochlore lattice takes the form 
\begin{align}
      \mathcal{S}^{\alpha\gamma}_{\mu\nu}=\left\langle S_{\mu}^\alpha (\bm{q}) S_{\nu}^\gamma(-\bm{q})\right\rangle ,
\end{align}
where $\mu,\nu$ label the sublattices and $\alpha,\beta$ label the spin components. Using this correlation function, we define the unpolarized neutron-structure factor 
\begin{align}
 \mathcal{S}_\perp(\bm{q}) &=&\sum_{\alpha,\beta}\sum_{\mu,\nu}\left(\delta_{\alpha,\beta} -\hat{\bm{q}}^\alpha \hat{\bm{q}}^\beta\right)\left\langle S_{\mu}^\alpha (\bm{q}) S_{\nu}^\beta(-\bm{q})\right\rangle,\label{eq:unpolarized}
\end{align}
where the factor in parenthesis projects out the component of the spins along the incident neutron propagation vector $\bm q$. Beyond the unpolarized neutron structure factor, we consider the polarized neutron structure factor $\mathcal{S}_\perp(\bm{q})$ in two channels, the non-spin-flip (NSF) and the spin-flip (SF) channels, defined in terms of the incident neutrons' polarization $\hat{Z}_{\textrm N}$~\cite{Chung_flatband} as
\begin{eqnarray}
 \mathcal{S}_\perp^{\mathrm{NSF}}(\bm{q})&=&\sum_{\alpha,\beta}\sum_{\mu,\nu}\left(\hat{z}_{\textrm N}
^\alpha \hat{z}_{\textrm N}
^\beta\right)\left\langle S_\mu^\alpha S_\nu^\beta \right\rangle,\label{eq:polarized_NSF}\\
\mathcal{S}_\perp^{\mathrm{SF}}(\bm{q})&=&\mathcal{S}_\perp (\bm{q})-\mathcal{S}_\perp^{\mathrm{NSF}}(\bm{q}).\label{eq:polarized_SF}
\end{eqnarray}
For the neutron scattering observed in the $[hh\ell]$ ($[hk0]$) plane, the incident polarization direction is defined as $\hat{Z}_{\textrm N}=[1\bar{1}0]$ ($[001]$).
\\

{\it Setup of pf-FRG calculations.---}
For the numerical implementation of pf-FRG, we extended the \texttt{PFFRGSolver.jl} Julia package \cite{pffrgsolver}, which provides state-of-the-art adaptive integration routines for solving the pf-FRG flow equations. In all calculations, the four-point vertex -- which depends on three independent frequency arguments due to energy conservation -- is discretized on a $35 \times 30 \times 30$ frequency grid, employing the efficient asymptotic parametrization in terms of one bosonic and two fermionic frequencies introduced in Ref.~\cite{wentzell2020}. Within this setup, we compute the spin-spin correlations defined in Eq.~\eqref{eq:correlations} throughout the $D_a$-$D_b$ plane down to cutoff scales of $\Lambda = 0.01J$, below which the numerical integration often becomes unstable. Based on these results, we construct the quantum phase diagram shown in Fig.~\ref{fig:quantum-phase-diagram}(b). Below, we outline the criteria used to determine the phase boundaries.

As a first step, we distinguish between dipolar magnetic order, i.e.\ order characterized by an order parameter linear in the spin operators, and quantum paramagnetic phases. Within pf-FRG, dipolar order is signaled by a divergence (or ``flow breakdown'') in selected components of the susceptibility $\chi^{\Lambda,ab}$, whereas paramagnetic phases are identified by the absence of such a breakdown. We note that numerical approximations required to solve the flow equations can soften or smear divergences, rendering the precise identification of flow breakdowns challenging. This is particularly relevant for the present model, where ordering transitions are expected to occur at very low energy scales of order $\sim 0.01J$, a regime in which the pf-FRG flow becomes increasingly difficult to resolve. In addition, the low symmetry of the breathing pyrochlore lattice restricts the accessible truncation lengths to relatively small values ($L \leq 5$).

\begin{figure}
    \centering
    \includegraphics[width=.95\textwidth]{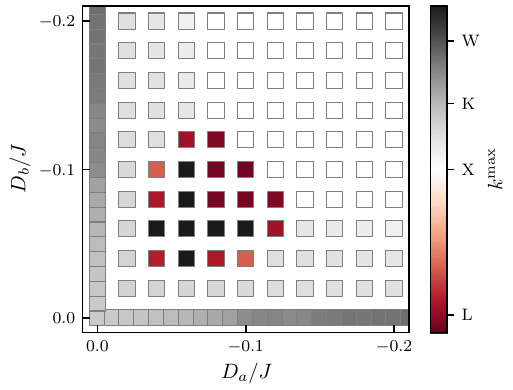}
    \caption{\textbf{Evolution of the momentum of maximal spectral weight in the quantum spin-spin correlations.}
    Shown is the magnitude of the momentum $\bm{k}^{\mathrm{max}}$ at which the static spin-spin correlation  $\chi^{zz}(\bm{k})$ attains its maximum in the low-cutoff limit.}
    \label{fig:qmaxs}
\end{figure}

Taking these limitations into account -- and noting that pf-FRG has a known tendency to overestimate the extent of paramagnetic regions -- we adopt a conservative criterion that favors the identification of ordered phases. Motivated by the behavior along the $D_b=0$ line, see Fig.~\ref{fig:quantum pinch-points}, where the onset of order is accompanied by a pronounced increase in both the absolute magnitude of $\chi^{zz}$ and its dependence on the truncation length $L$, we classify a phase as magnetically ordered if the maximum of the susceptibility flow, $\chi^{zz}(\bm{k}^{\mathrm{max}})$, increases by at least one order of magnitude between truncation lengths $L=3$ and $L=5$. This criterion is encoded in the color scale and solid phase boundaries shown in Fig.~\ref{fig:quantum-phase-diagram}(b). We emphasize that this choice is slightly more stringent, i.e.\ biased toward identifying order, than criteria commonly employed in earlier pf-FRG studies.

In regions exhibiting dipolar order, we further distinguish between different ordered phases by analyzing the momentum dependence of the susceptibility $\chi^{zz}(\bm{k})$ and comparing neutron scattering structure factors with their classical counterparts. The evolution of the momentum $\bm{k}^{\mathrm{max}}$ at which $\chi^{zz}(\bm{k})$ is maximal across the phase diagram is shown in Fig.~\ref{fig:qmaxs}. For large $D_a,D_b<0$, the maximum occurs at $\bm{k}^{\mathrm{max}}=\bm{X}$, in agreement with $\Gamma_5$ order. Along the $D_a=0$ ($D_b=0$) axes, the peak shifts toward $\bm{k}^{\mathrm{max}}=\bm{W}$. In the incommensurate spiral (ICS) phase, $\bm{k}^{\mathrm{max}}$ varies continuously and does not coincide with any high-symmetry point of the Brillouin zone.

In quantum paramagnetic regions, we estimate the extent of the rank-1 and rank-2 $U(1)$ spin-liquid phases by inspecting the structure factor for the presence of two- or fourfold pinch-point singularities, as discussed in the main text. We stress that this procedure is not quantitative and relies on visual inspection of the correlation functions. Consequently, the corresponding phase boundaries, indicated by dashed lines in Fig.~\ref{fig:quantum-phase-diagram}(b), should be interpreted as approximate.\\

{\it Classical Monte Carlo simulations.---} 
To study a system described by the classical model given in Eq.~\eqref{eq:H+DM}, we employed a classical Monte Carlo simulation for a system defined with an FCC unit cell composed of $4L^3$ spins with $L=10$ where we consider Heisenberg spins with a length constraint $|\bm S|=1$. In our simulations, we used a combination of local Gaussian spin updates~\cite{Alzate-Cardona_2019}, and an over-relaxation update~\cite{ZhitomirskyPRL2012,Creutz}. To further improve the thermal averaging of our simulations, we averaged the results obtained from $10$ independent Monte Carlo runs. We performed $8\times10^4$ thermalization sweeps and $ 2\times10^5$ measurement sweeps where we measured various thermodynamic quantities of interest. 

\begin{figure}[h!]
    \centering
      \begin{overpic}[width=.95\columnwidth]{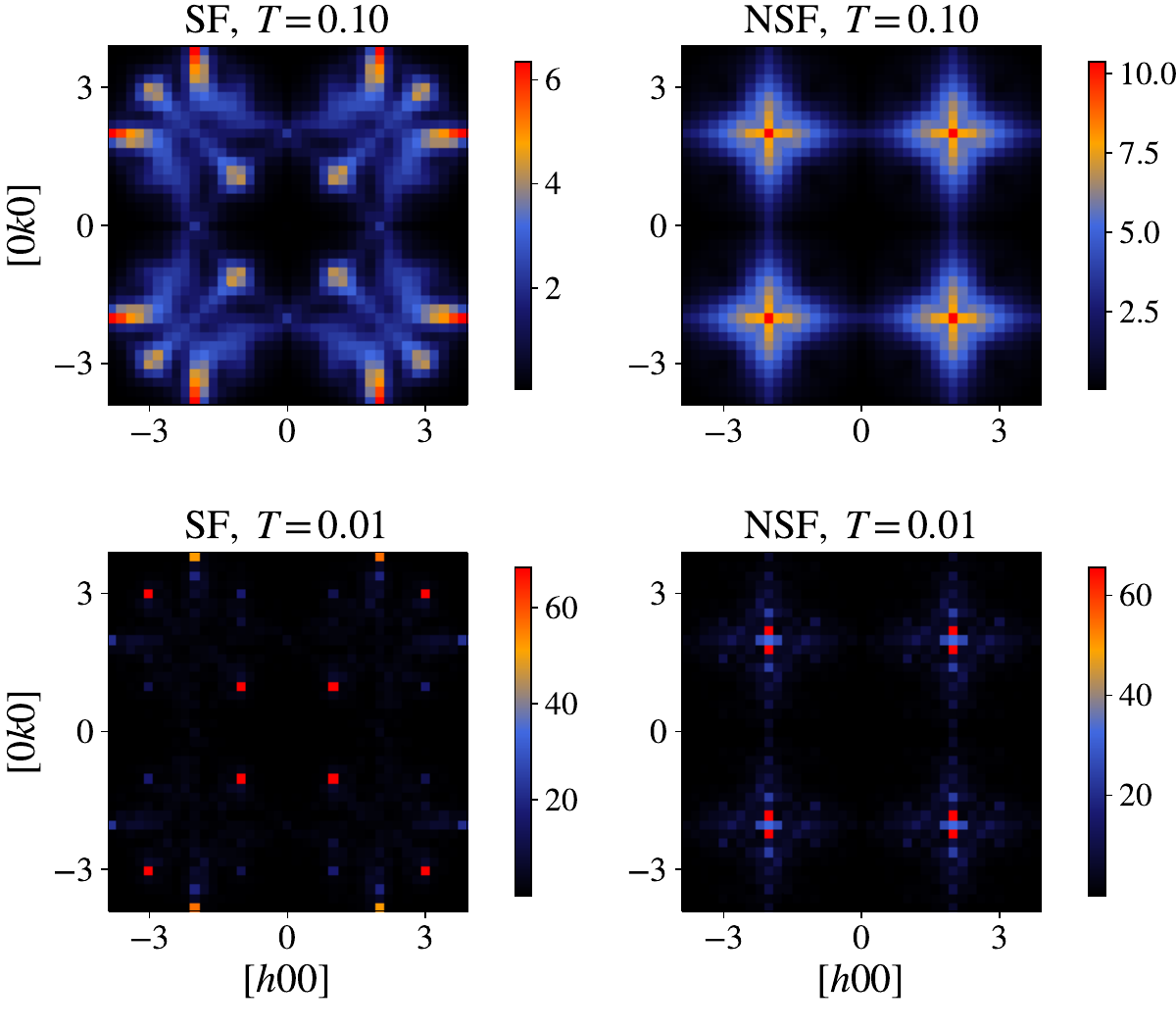}
        \put(5,86){(a)}
        \put(55,86){(b)}
         \put(5,43){(c)}
        \put(55,43){(d)}
    \end{overpic}
\caption{{\bf Polarized neutron structure factors} obtained via classical Monte Carlo with the same interactions parameters used to obtain the specific heat curve in Fig.~\ref{fig:classical_pinch-points}(f) in the main text. Panels (a) and (b) show the spin-flip (SF) and non-spin-flip (NSF) channels of the polarized neutron scattering for a temperature above the transition temperature $T_c$. Panels (c) and (d) illustrate the equivalent measurements as those in panels (a) and (b), respectively, for a temperature below the transition temperature $T_c$. }\label{fig:CMC_structures}
\end{figure}
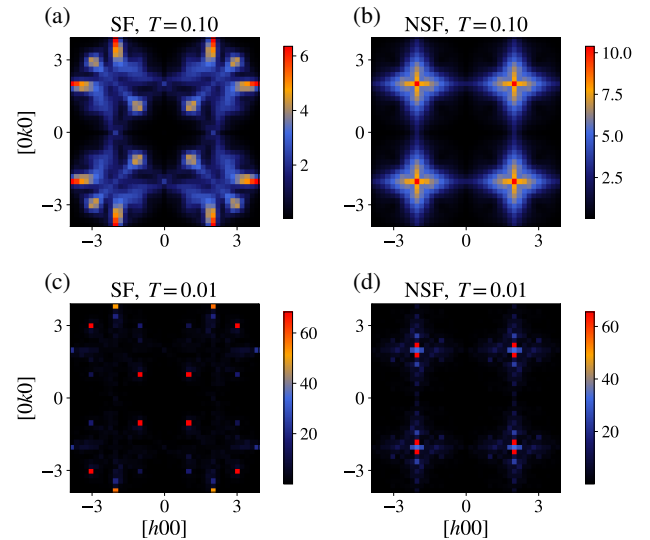

As reported in~\cite{Yan-2020,pollet_ML_R2}, for systems with sets of parameter interactions along the $D_b=0\ (D_a=0)$ line, our simulations below the critical temperature $T_c$ become increasingly complex, where the system might become trapped in a local minimum~\footnote{Reference~\cite{pollet_ML_R2} demonstrated that a better thermal average can be obtained by supplementing the classical Monte Carlo simulations with additional information regarding the configuration-space low-energy landscape provided by a Support-Vector Machine.}. Such behavior is not observed above the critical temperature $T_c$, where a classical rank-2 $U(1)$ spin liquid is realized. In Fig.~\ref{fig:CMC_structures}, we illustrate the polarized neutron scattering obtained for such a system at a temperature above and below the transition temperature $T_c$. Above the critical temperature $T_c$, the polarized neutron structure factor obtained via classical Monte Carlo closely resembles the predictions obtained via SCGA, see Fig.~\ref{fig:classical_pinch-points}(c) and (d) and Fig.~\ref{fig:CMC_structures}(a) and (b). Below the critical temperatures, the system undergoes a symmetry-breaking transition resulting in the observation of high-intensity peaks associated with a $\bm q=\bm W$ order. 

\clearpage

\begin{thebibliography}{53}
\bibitem [{fn_()}]{fn_rank2width}%
  \BibitemOpen
  \href@noop {} {}\bibinfo {note} {{For this simplified model the low-energy
  bands of the interaction matrix $J_{ij}^{\alpha\beta}$ are completely flat.
  It is however worth mentioning that this feature is kept even when the
  Heisenberg coupling is positive but not equal in both tetrahedra, i.e.
  $J_a\neq J_b$~\cite{Benton-2015}}}\BibitemShut {NoStop}%
\end{thebibliography}

%

\clearpage

\newcommand{\addpage}[1] {
\begin{figure*}
  \includegraphics[width=8.5in,page=#1]{Rank_2_BreathingPyro_DM_SM.pdf}
\end{figure*}
}
\addtolength{\oddsidemargin}{-0.75in}
\addtolength{\evensidemargin}{-0.75in}
\addtolength{\topmargin}{-0.725in}
\addpage{1}
\addpage{2}
\addpage{3}

\end{document}